\documentclass[11pt,twoside]{article}  
\usepackage{adassconf}

\begin{document}   

\paperID{P02.12}

\title{The Skysoft Project}

\author{C.\ Baffa, E.\ Giani, A.\ Checcucci}
\affil{INAF - Osservatorio Astrofisico di Arcetri, Largo E.Fermi 5,
       50125 Firenze, Italy}
\affil{and IRA-CNR, Largo E.Fermi 5, 50125 Firenze, Italy}

\contact{C.Baffa}
\email{baffa@arcetri.astro.it}

\paindex{Baffa, C.}
\aindex{Giani, E.}     
\aindex{Checcucci, A.}

\authormark{Baffa, Giani \& Checcucci}


\begin{abstract}          
We present the Skysoft project. Skysoft is a YAASD
(Yet Another Astronomical Software Directory), but with a different overall
approach.

To be useful, Skysoft needs to be a long--lived project, setting little
pressure for maintenance, imposing a very low nuisance level to the
developers community, and requiring a low maintenance cost. Our aim is to
design Skysoft as a community--supported directory, to which everyone can
contribute, both developers and end--users.

\end{abstract}

\section{Introduction}

\htmladdnormallinkfoot{Skysoft}{http://www.skysoft.org} is 
an astronomical software directory, but with a
peculiar overall approach. Our choice is to design this site as a
community supported directory. All people can contribute, with
software news, user's views, comments and bugs/refuses notifications.
Developers should post a brief description of their product, with the
classification which can ease the search and retrieval of software
projects.

\section{What is Skysoft site about?}

Skysoft is designed  as a community supported directory. The
idea is that, like a chat session, content remains timely because of
frequent user interaction. Users and developers are invited to contribute
with software news, user's views, comments and bugs notifications. This
site is designed so that software developers acquire more visibility because
of astronomical context.  Developers can post a brief description of their
product, with the classification which can ease the search and retrieval of
software projects.

\section{Why not a traditional site?}

Traditional sites, such as
\htmladdnormallinkfoot{ASDS}{http://asds.stsci.edu} and
\htmladdnormallinkfoot{ASCL}{http://ascl.net} are valuable, and
widely used. But we think they are most useful in the standard context
of mainstream data analysis and reduction, where ten-twelve
applications do 95\% of the work, and remain there for many years. It is
difficult for traditional sites to easily accommodate new ideas, new
approaches for less-used telescopes+instruments. For instance, during
data mining, we have found several interesting approaches to the same
specialized problem (not addressed by mainstream tools), but which were
rewritten over a decade by different groups, each without knowledge of
others' efforts.

We propose a faster and more flexible approach as a complement to
traditional sites. Also we do not need to have all the expertise in all the
fields Skysoft covers: it is enough that such expertise resides in the
users' community!

\section{What is the Skysoft approach and what are its possible advantages?}

Skysoft is intended to be built by the community which uses it! If you
think that Skysoft lacks some information you deem useful, just add it!
Many others can benefit by your (minimum) effort! Plus you gain publicity
for your work!

Our aim is to build a site useful for astronomers and instrument
developers, and to make this utility widely available, easy to use, and
up-to-date with the latest developments. We cannot cope with the enormous
amount of information and expertise needed. But the community as a whole
has all the necessary competence! If we all share our 2 cents of
informations, we will build a site more useful for everybody.

We started with a small amount of software we found in the net, just to
boostrap the site. The selection was rather arbitrary, based on our own
knowledge. Obviously, we have missed important informations: please add it
and help us to improve the site!  We are ready to add a newsletter, an
event calendar, some discussion lists, and more.

\section{So do I absolutely need Skysoft?}

No. Everything which is available at Skysoft site can be found using other
Astronomical software collections, or asking Google, or asking some
colleagues and friends. But wait, my last Google interrogation returned
112000 documents! A more specialized site can help  speed up things
significantly!! We aim to be a first choice in search process! 

\section{Present structure}

The Skysoft Software Database has been structured as a tree. This structure
is wrong in our view: the software has a structure by far too complex to be
represented in such a bidimensional way. Consideration of Database
implementation and ease of consultation led us to strip down such a
complexity.

Here there is the database structure at the time of last modification. By
the dynamic nature of Skysoft approach, this can be outdated.

\begin{center}

{\scriptsize
\parbox{3in}{
\begin{itemize}

 \item Data reduction and Handling
       \subitem Large general packages
                \subsubitem add-on for large packages
       \subitem Data Handling and Display
       \subitem Data Conversion
       \subitem Statistical packages
       \subitem Utilities
\smallskip
 \item Data Archiving
       \subitem Databases
       \subitem Query Languages
       \subitem Structure Organizer
\smallskip
 \item Astronomical Tools
       \subitem Large Tools collections
       \subitem Sky maps
       \subitem Coordinate conversion
       \subitem Celestial Ephemeris
       \subitem Observations planning
       \subitem Physical simulation and modelling
\smallskip
 \item Developer's Tools
       \subitem  General Tools
       \subitem  Device Drivers
       \subitem  Human Interfaces
       \subitem  Data Handling libraries
       \subitem  Data Archiving libraries
\smallskip
 \item Management
       \subitem  Time allocation
       \subitem  Telescope pointing
       \subitem  Project management
\smallskip
 \item Didactic Tools
       \subitem  Digital Orreries
       \subitem  Learning tools
       \subitem  Amateur's tools

\end{itemize}
}
}
\end{center}

\begin{figure}
\plotfiddle{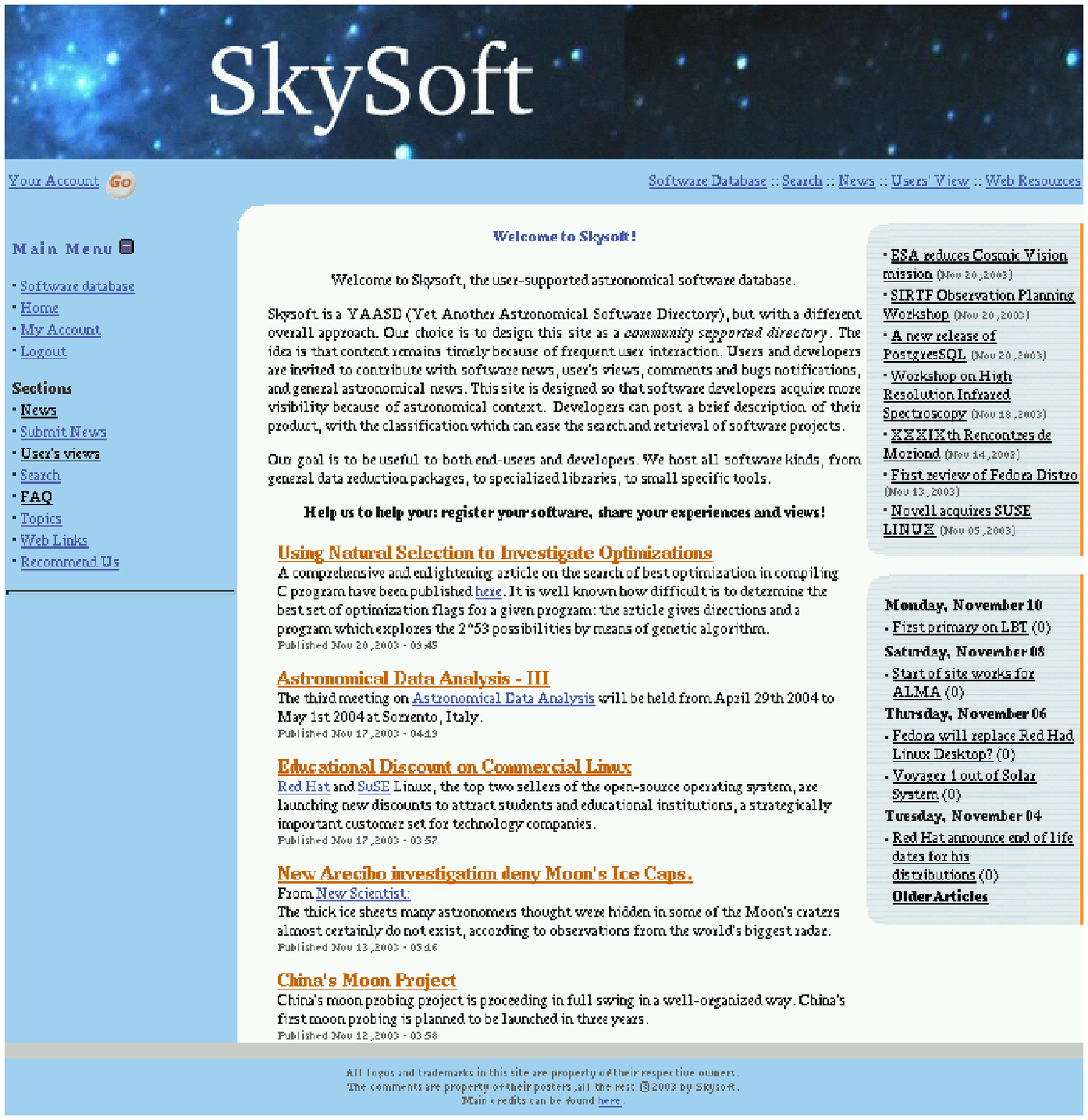}{18cm}{0}{80}{90}{-250}{-130}
\caption{The Skysoft front page}
\end{figure}

\acknowledgments
Skysoft lives by the collaboration of really many people:
thanks to all our guests, collaborator, maintainers and data miners. 
Thanks to G.Calculli and F.Giovannini for their advise.
Founding and support has been provided by:
INAF/IRA and Arcetri Astrophysical Observatory.

\end{document}